\documentclass[%
 aip, floatfix
 amsmath,amssymb,
 reprint,%
]{revtex4-1}
\usepackage{graphicx}% Include figure files
\usepackage{dcolumn}% Align table columns on decimal point
\usepackage{bm}% bold math
\usepackage{epsfig,wrapfig}
\usepackage{epstopdf} % for pdflatex
\usepackage{fancyhdr}
\usepackage{indentfirst}
\usepackage{titlesec}
\usepackage{amsmath}
\usepackage{color}
\usepackage{setspace}
\usepackage{gensymb}
\usepackage{textcomp}
\usepackage[utf8]{inputenc}
\usepackage[T1]{fontenc}
\usepackage{mathptmx}
\usepackage{amsmath}
\usepackage[rightcaption]{sidecap}
\usepackage{ulem}
\begin{document}

\title
{Acoustic imaging by three-dimensional acoustic Luneburg meta-lens with lattice columns}

\author{Jung-Woo Kim}
\affiliation{School of Mechanical Engineering, Gwangju Institute of Science and Technology, 123 Cheomdangwagi-ro, Buk-gu, Gwangju 61005, Republic of Korea}

\author{Seong-Jin Lee}
\affiliation{Division of Marine Engineering, Mokpo National Maritime University, 91 Haeyangdaehak-ro, Mokpo 58628, Republic of Korea}

\author{Jun-Yeong Jo}
\affiliation{Division of Marine Engineering, Mokpo National Maritime University, 91 Haeyangdaehak-ro, Mokpo 58628, Republic of Korea}

\author{Semyung Wang}
\affiliation{School of Mechanical Engineering, Gwangju Institute of Science and Technology, 123 Cheomdangwagi-ro, Buk-gu, Gwangju 61005, Republic of Korea}

\author{Sang-Hoon Kim}
\email{shkim@mmu.ac.kr, Corresponding author}
\affiliation{Division of Marine Engineering, Mokpo National Maritime University, 91 Haeyangdaehak-ro, Mokpo 58628, Republic of Korea}

\begin{abstract}
A three-dimensional acoustic Luneburg meta-lens has the advantage of refracting sound waves for all incident angles and focusing higher sound pressure compared to a two-dimensional lens.
The lens made of plastic with a diameter of 120 mm was designed with thousands of lattice column-shaped meta-atoms to maintain its three-dimensional shape.
The lens's three-dimensional focusing performance and acoustic imaging were simulated and measured at the frequency range of 5 kHz $\sim$ 20 kHz.
The omnidirectional property was confirmed by rotating the lens to change the incident angle and measuring the sound pressure.
The development of these spherical Luneburg meta-lenses is expected to improve the performance of devices that require acoustic focusing.
\end{abstract}
\maketitle

%%%%%%%%%%%%%%%%%%%%%%%%%%%%%%%%%%%%%%%%%%%%%%%%%%%%%%%%%%%%%%%%%
%\section{Introduction}
%%%%%%%%%%%%%%%%%%%%%%%%%%%%%%%%%%%%%%%%%%%%%%%%%%%%%%%%%%%%%%%%%

The Luneburg lens, which was proposed as a model of the wave transformation theory by L{\"u}neburg in the 1940s,
is a type of gradient index (GRIN) lens that can focus incoming waves on the opposite side of the lens without aberration.\cite{lune}
Some theoretical and experimental validation of the lens was conducted in the 1960s,\cite{peel,gund} and the lenses with metamaterial structures were realized
in various methods and shapes in the 21st century.\cite{chen,dock,zenn,ma}
Its high gain and spherical symmetry advantages, design optimization,\cite{hoss,hsi}
printed circuit board fabrication technique,\cite{carl,yuan}
quasi-conformal transformation optics,\cite{soum} and metamaterial structures\cite{abda,mark,dani,bohu}
were used to apply the lenses to antennas.
These studies have been mainly conducted in the electromagnetic and optical fields,
but attempts have also been made to apply them to sound waves.

With the advent of metamaterials, research on the implementation of acoustic focusing\cite{torr1, peng,zigo,yong,torr2,jahd,qian}
was conducted to overcome the limitations of existing materials by controlling the waves, and the Luneburg lens was no exception.
Kim realized the performance of a two-dimensional (2D) acoustic Luneburg meta-lens in the audible frequency range.\cite{kim1,kim2}
Moreover, many methods and shapes have been proposed to achieve this.
An acoustic Luneburg lens composed of orifice-type unit cells\cite{park} and a 2D flattened lens designed using a quasi-conformal mapping method\cite{dong,yu} has been presented.
Anisotropic acoustic Luneburg lenses such as Luneburg-fisheye combined\cite{fang} and Luneburg retroreflector\cite{fu-cummer} were introduced.

Research on the acoustic Luneburg meta-lenses has expanded to a three-dimensional (3D) shape because the 3D lenses have the advantage of focusing incident waves in all directions and achieving a higher sound pressure level (SPL) than the 2D lenses.
Fabricated 2D and 3D lenses, and the performance of the 2D lens in the ultrasound frequency range were studied by Xie et al.\cite{xie-cummer}
After that, Zhao et al. reported ultrasound beam steering using 2D and 3D flattened acoustic metamaterial Luneburg lenses at 40 kHz.\cite{zhao}
Despite these pioneering studies, frequency-dependent focusing performance and acoustic imaging, consistent with 3D numerical simulation and experimental results for full 3D lenses, has not been reported.

In this paper, a 3D acoustic Luneburg meta-lens, or an acoustic Luneburg ball, composed of lattice column-shaped meta-atoms to maintain its three-dimensional shape was designed and fabricated, and its focusing performance was confirmed using COMSOL Multiphysics.
Acoustic imaging was measured at the frequency range of 5 kHz $\sim$ 20 kHz using the transducer, microphone, three-axis linear stage, and control module.
It was then demonstrated to have omnidirectional property by rotating the lens to change the incident angle and measuring the sound pressure around the lens.

%%%%%%%%%%%%%%%%%%%%%%%%%%%%%%%%%%%%%%%%%%%%%%%%%%%%%%%%%%%%%%%%%
%\section{Design and Simulation}
%%%%%%%%%%%%%%%%%%%%%%%%%%%%%%%%%%%%%%%%%%%%%%%%%%%%%%%%%%%%%%%%%

\begin{SCfigure*}[1]
\centering
\includegraphics[width=125mm]{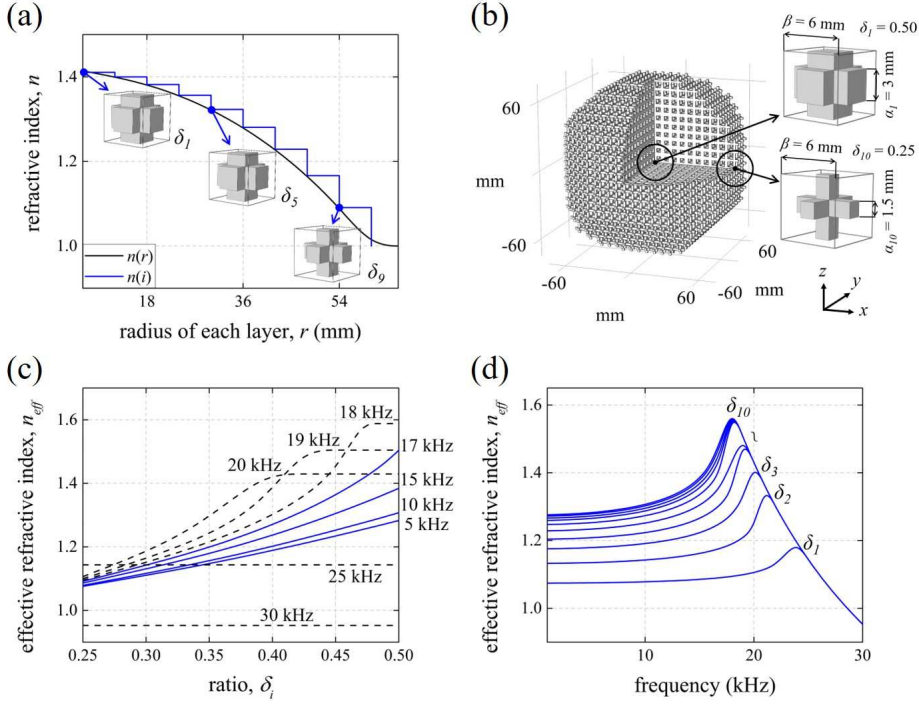}
\caption{Refractive index and structure of the 3D acoustic Luneburg meta-lens. (a) Refractive index $n$ by the radius of each layer $r$. (b) Luneburg ball and its lattice column unit cells. (c) Ratio $\delta_i$- and (d) frequency-dependent effective refractive index $n_{eff}$.}
\label{Fig.1}
\end{SCfigure*}

The refractive index (RI) formula of the Luneburg lens is $n(r)=n_o\sqrt{2-(r/R)^2}$, where $n_o$ is the RI of the surrounding medium, $R$ is the radius of the lens, and $r$ is the distance from the center of the lens.
It is possible to discretize the volume as many concentric layers due to the spherical symmetry.
If the lens is divided into N layers, the RI is newly defined as
%Eq. 1
\begin{equation}
n(i)=\sqrt{2-\left(\frac{i}{N}\right)^2},
\label{Eq.1}
\end{equation}
where $i$ = 0, 1, 2, ..., ($N$-1) and $N$ = 10 is chosen in the design.
The lattice columns' sizes increase toward the center of the lens, and the RI changes for each layer, as shown in Fig.~\ref{Fig.1}(a).
$\alpha_i$ and $\beta$ are the dimensions of the meta-atoms and unit cells.
The radius of the lens $R$ is 60 mm, and the length of the unit cell $\beta$ is 6 mm.
$\alpha_i$ is the length of one side of the $i^{th}$ meta-atoms and varies depending on each layers' position.
Therefore, the ratio is set to $0<\delta_i=\alpha_i/\beta<1$.
It can be seen that the RI is derived as $n(\delta_i)=1/\sqrt{1-3\delta_i^2+2\delta_i^3}$ and the index varies according to the size of the meta-atoms in Fig.~\ref{Fig.1}(a).
The lens is designed in a form in which 19 layers are stacked while maintaining the 3D shape in Fig.~\ref{Fig.1}(b).

%%%%%%%%%%%%%%%%%%%%%%%%%%%%%       Fig. 1      %%%%%%%%%%%%%%%%%%%%%%%%%%%%

%%%%%%%%%%%%%%%%%%%%%%%%%%%%%       Fig. 2      %%%%%%%%%%%%%%%%%%%%%%%%%%%%
\begin{SCfigure*}[1]
\centering
\includegraphics[width=130mm]{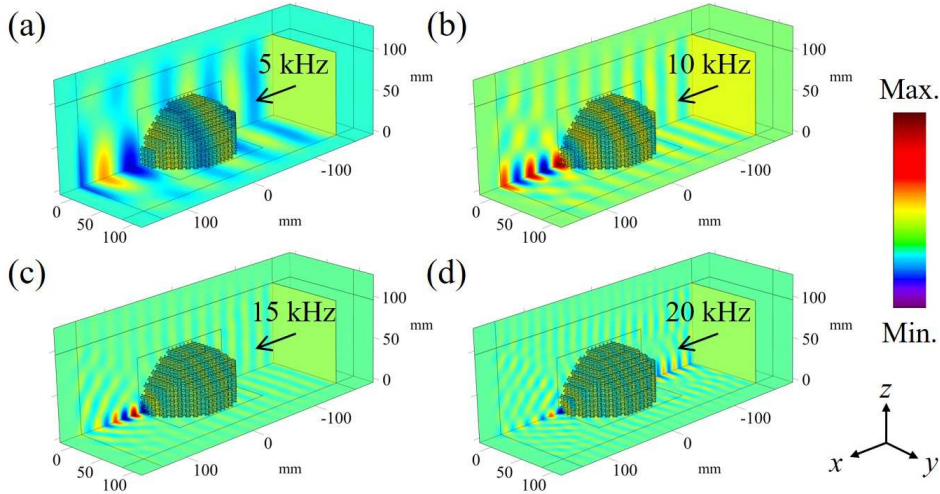}
\caption{Numerical simulation results for the frequency-dependent focusing performance at (a) 5 kHz, (b) 10 kHz, (c) 15 kHz, and (d) 20 kHz.}
\label{Fig.2}
\end{SCfigure*}
%%%%%%%%%%%%%%%%%%%%%%%%%%%%%%%%%%%%%%%%%%%%%%%%%%%%%%%%%%%%%%%%%%%%%%%%%%%

The index was also calculated to retrieve effective material properties of the acoustic metamaterials.
It was simulated to obtain the reflection $R$ and transmission coefficients $T$ using COMSOL Multiphysics, a finite element analysis.
The formulas for the effective refractive index $n_{eff}$ and acoustic impedance $\xi$ derived from the retrieval method are as follows.\cite{fokin}
%%%%%%%%%%%%%%%%%%%%%%%%%%%  Eq. 2  %%%%%%%%%%%%%%%%%%%%%%%%%%%%%%%%%%%%%%
\begin{equation}
n_{eff} = \frac{1}{kd}\left\{2 \pi m \pm cos^{-1}\left(\frac{1-(R^2-T^2)}{2T}\right)\right\},
\label{Eq.2}
\end{equation}
%%%%%%%%%%%%%%%%%%%%%%%%%%%  Eq. 3  %%%%%%%%%%%%%%%%%%%%%%%%%%%%%%%%%%%%%%
\begin{equation}
\xi = {\pm}\sqrt{\frac{\left(1+R\right)^2-T^2}{\left(1-R\right)^2-T^2}},
\label{Eq.3}
\end{equation}
%%%%%%%%%%%%%%%%%%%%%%%%%%%%%%%%%%%%%%%%%%%%%%%%%%%%%%%%%%%%%%%%%%%%%%%%%%
where $k$ is the wavenumber, $d$ is the slab thickness, and $m$ is the branch number of $cos^{-1}$ function.
The relationship between the effective refractive index and ratio and the frequency-dependent index
are plotted in Figs.~\ref{Fig.1}(c) and~\ref{Fig.1}(d).
It was shown that the index does not gradually change above 18 kHz.
Therefore, the lens has a frequency-dependent property, and the focusing performance depends on the frequency.
From the two figures, it can be predicted that the applicable frequency range is up to approximately 17 kHz.

The simulated focusing performance of the polyamide plastic lens ($\rho$ = 1,000 kg/m$^3$ and $c$ = 2,200 m/s) in the air ($\rho$ = 1.2 kg/m$^3$ and $c$ = 343 m/s) is shown in Fig.~\ref{Fig.2} expressed as sound pressure.
The acoustics module was used in the frequency domain, solving the Helmholtz equation to analyze its acoustics effects and check the performance by frequency.
As a result of the simulation, the sound waves are focused from 5 to 15 kHz, but not at 20 kHz.
In general, the size of the unit cell is a subwavelength dimension, so that the lens does not work at the frequency of 20 kHz ($\lambda/4$ = 4.875 mm).
In addition, this reason can also be confirmed in Figs.~\ref{Fig.1}(c) and~\ref{Fig.1}(d).
Therefore, the simulation and applicable frequency range of the lens matches.

%%%%%%%%%%%%%%%%%%%%%%%%%%%%%%%%%%%%%%%%%%%%%%%%%%%%%%%%%%%%%%%%%%%%%%%%
%\section{Experiments}
%%%%%%%%%%%%%%%%%%%%%%%%%%%%%%%%%%%%%%%%%%%%%%%%%%%%%%%%%%%%%%%%%%%%%%%%

%%%%%%%%%%%%%%%%%%%%%%%%%%%%      Fig. 3     %%%%%%%%%%%%%%%%%%%%%%%%%%%
\begin{figure}[t!]
\includegraphics[width=85mm]{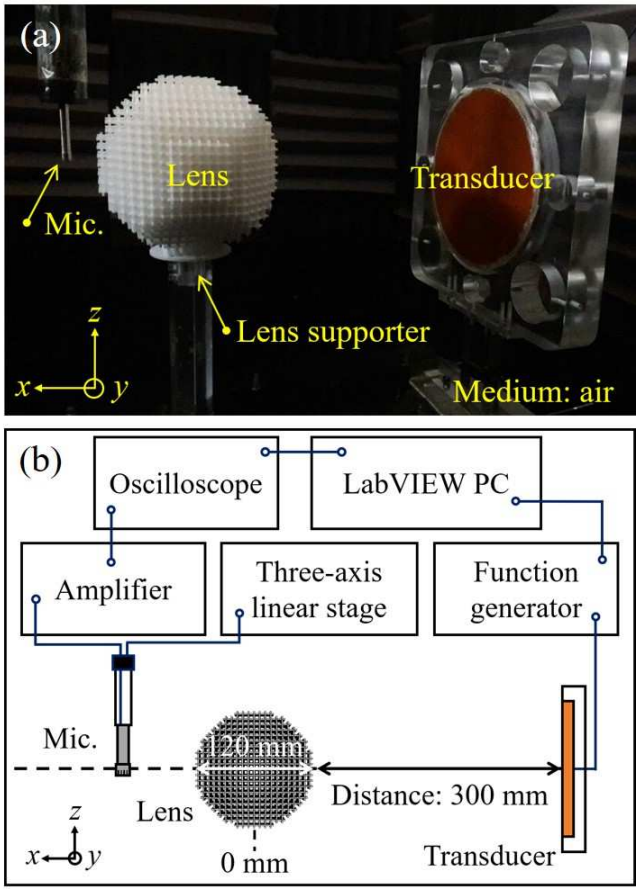}
\caption{(a) Photograph of the experimental setup. (b) Schematic diagram for the measurement.}
\label{Fig.3}
\end{figure}

%%%%%%%%%%%%%%%%%%%%%%%%%%%%       Fig. 4    %%%%%%%%%%%%%%%%%%%%%%%%%%%
\begin{figure}[t!]
\includegraphics[width=80mm]{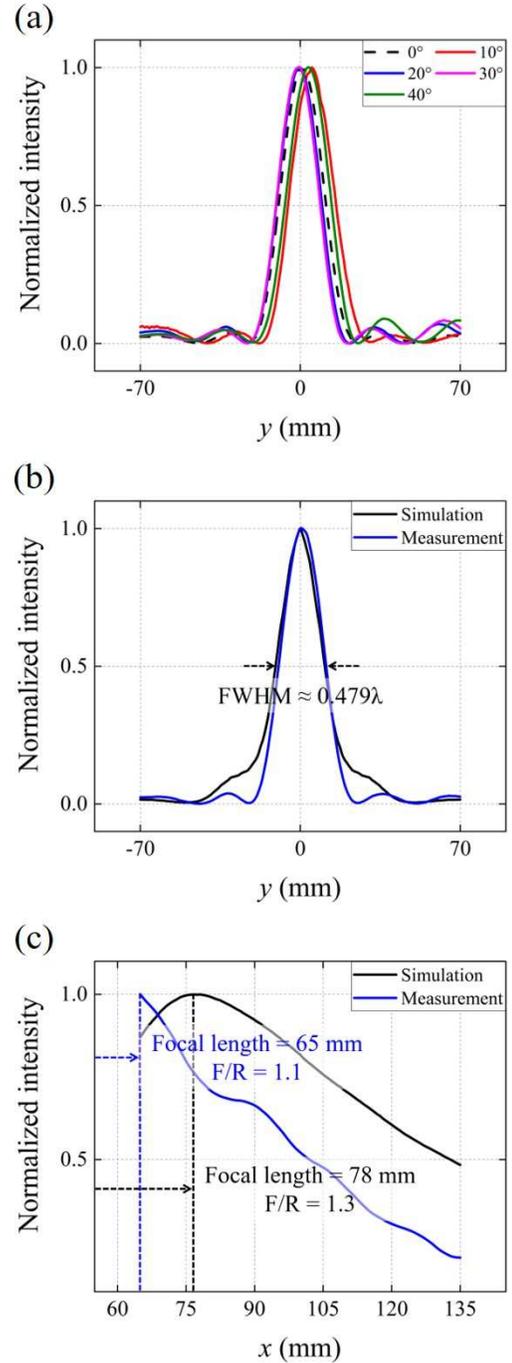}
\caption{(a) Focusing performance at a frequency of 10 kHz depending on the incident angles (0$\degree$, 10$\degree$, 20$\degree$, 30$\degree$, and 40$\degree$). Focusing performance at a frequency of 10 kHz on (b) y- and (c) x-axis of the simulation and measurement when the incident angle is 0$\degree$.}
\label{Fig.4}
\end{figure}

%%%%%%%%%%%%%%%%%%%%%%%%%%%%%       Fig. 5     %%%%%%%%%%%%%%%%%%%%%%%%%%%%
\begin{figure*}
\centering
\includegraphics[scale=0.5]{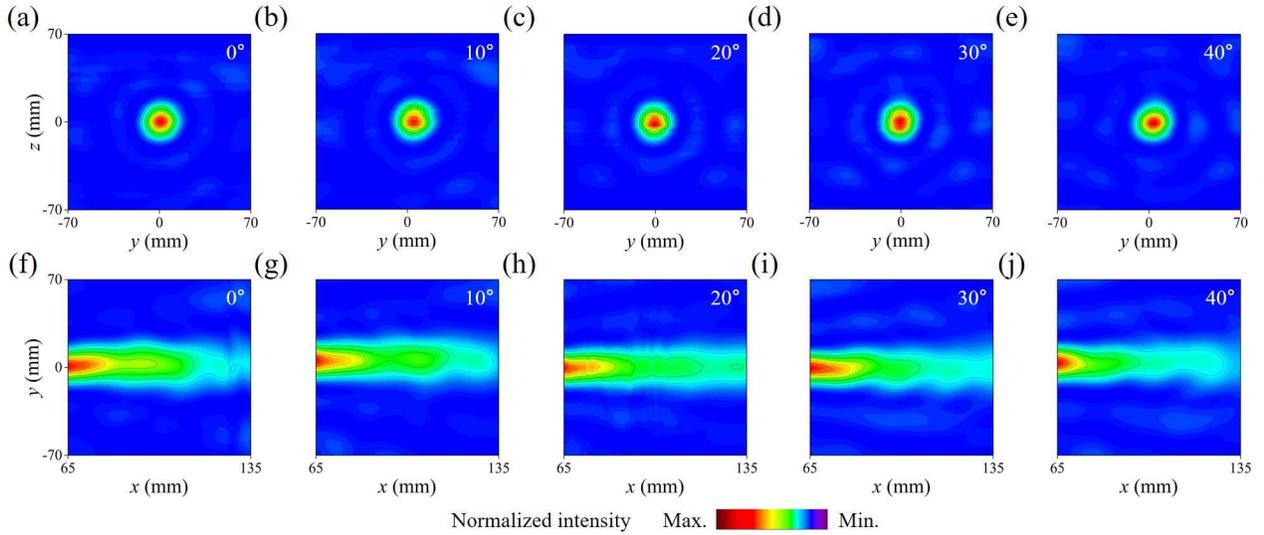}
\caption{Experimental results of acoustic imaging at a frequency of 10 kHz. (a)-(e) Acoustic imaging in the yz-plane with the incident angles of 0$\degree$, 10$\degree$, 20$\degree$, 30$\degree$, and 40$\degree$. (f)-(j): Acoustic imaging in the xy-plane with the incident angles are 0$\degree$, 10$\degree$, 20$\degree$, 30$\degree$, and 40$\degree$. These results are normalized intensity data.}
\label{Fig.5}
\end{figure*}
%%%%%%%%%%%%%%%%%%%%%%%%%%%%%%%%%%%%%%%%%%%%%%%%%%%%%%%%%%%%%%%%%%%%%%%%%%

The lens was fabricated using a selective-laser-sintering 3D printer.
The experimental setup was configured to validate the performance of the fabricated lens in Fig.~\ref{Fig.3}.
The reflection of sound waves was minimized by surrounding the measurement space with a sound-absorbing material.
The supporter was connected to the plastic structure so that the lens is completely fixed to the bottom.
When sound waves move from the transducer to the lens, the sound pressure was measured around the lens while moving in the three-axis direction of a microphone connected to the three-axis linear stage.
The transducer generated acoustic pulse signals, which had a signal amplitude of 2 V, burst count of 25, burst period of 50 ms, and wait time of 500 ms.
The pressure was measured while rotating the lens (10$\degree$, 20$\degree$, 30$\degree$, and 40$\degree$) to demonstrate the omnidirectional property.
Acoustic imaging was measured by scanning the yz- (140 x 140 mm$^2$) and xy-planes (70 x 140 mm$^2$) of the lens's output area by 1 mm step size.
The SPL ($L_p$=20$log(p/p_o)$, $p$; sound pressure, $p_o=20$ $\mu$Pa; reference sound pressure in the air) recorded in the LabVIEW PC has been converted to the intensity ($I=p^2/Z_0$, $Z_0$ = 411.6 rayl; specific acoustic impedance in the air) to represent the full width at half maximum (FWHM) and focal length.

The rotational symmetry is shown in Fig.~\ref{Fig.4}(a) expressed as normalized intensity.
The comparison between numerical simulation and measurement results at a frequency of 10 kHz in the y- and x-axis is shown in Figs.~\ref{Fig.4}(b) and~\ref{Fig.4}(c), respectively.
The FWHM is approximately 16.426 mm (= 0.479$\lambda$), and the focal lengths are 78 mm (F/R = 1.3) and 65 mm (F/R = 1.1) in the simulation and measurement, respectively.
In principle, the focusing should be perfect (F/R = 1); however, it is a bit broad.
This is because the refractive indices of the ideal, designed, and fabricated lenses are different due to discretization into several layers and fabrication errors.
As a result, the focal length is larger than the radius of the lens.
Therefore, the focal point is formed exterior to the lens and not the interior.\cite{boyl1,boyl2}
Acoustic imaging expressed as normalized intensity for the yz- and xy-planes
at a frequency of 10 kHz was plotted in Fig.~\ref{Fig.5}.
The rotational symmetry was confirmed in the yz-plane (Figs.~\ref{Fig.5}(a)-\ref{Fig.5}(e)) and in the xy-plane (Figs.~\ref{Fig.5}(f)-~\ref{Fig.5}(j)).

%%%%%%%%%%%%%%%%%%%%%%%%%%%%%%%%%%%%%%%%%%%%%%%%%%%%%%%%%%%%%%%%%%%%%%%%%%
%\section{Summary}
%%%%%%%%%%%%%%%%%%%%%%%%%%%%%%%%%%%%%%%%%%%%%%%%%%%%%%%%%%%%%%%%%%%%%%%%%%%

The design, simulation, fabrication, and measurement of the 3D acoustic Luneburg meta-lens with a diameter of 120 mm were studied.
The meta-atoms were designed as lattice columns to realize a 3D lens discretized into 10 layers.
The frequency-dependent focusing performance was verified through the retrieval method and numerical simulation.
Its omnidirectional property was confirmed by changing the incident angle while rotating the lens.
The applicable frequency range was 5 to 17 kHz.
Acoustic imaging was measured in the yz- and xy-planes at a frequency of 10 kHz.
These focusing and beam steering functions can detect weak signals from all directions or increase imaging performance.
Therefore, this makes them suitable for sensing applications such as acoustic imaging systems, acoustic communication, non-destructive testing, and energy harvesting.\cite{miao,tol,jin,kim3,cheng}

\section*{ACKNOWLEDGEMENTS}
This work was mainly supported by the 2019 Undergraduate Research Program of the Korean Foundation for the Advancement of Science $\&$ Creativity, and partially by GIST Research Institute (GRI) grant funded by the GIST in 2020. The authors acknowledge helpful discussions with D. Lee and J. Rho from Pohang University of Science and Technology (POSTECH).

\section*{DATA AVAILABILITY}
The data that support the findings of this study are available from the corresponding author upon reasonable request.


\section*{REFERENCES}

\begin{thebibliography}{99}

\bibitem{lune}
R. K. L{\"u}neburg and M. Herzberger, {\itshape Mathematical Theory of Optics} (University of California Press, Berkeley, Los Angeles, 1964).

\bibitem{peel}
G. Peeler and H. Coleman, IRE Trans. Antennas Propag. \textbf{6}, 202 (1958).

\bibitem{gund}
L. C. Gunderson and G. T. Holmes, Appl. Opt. \textbf{7}, 801 (1968).

\bibitem{chen}
Q. Cheng, H. F. Ma, and T. J. Cui, Appl. Phys. Lett. \textbf{95}, 181901 (2009).

\bibitem{ma}
H. F. Ma and T. J. Cui, Nat. Commun. \textbf{1}, 124 (2010).

\bibitem{dock}
J. A. Dockrey, M. J. Lockyear, S. J. Berry, S. A. R. Horsley, J. R. Sambles, and A. P. Hibbins, Phys. Rev. B \textbf{87}, 125137 (2013).

\bibitem{zenn}
Y.-Y. Zhao, Y.-L. Zhang, M.-L. Zheng, X.-Z. Dong, X.-M. Duan, and Z.-S. Zhao, Laser Photonics Rev. \textbf{10}, 665 (2016).

\bibitem{hoss}
H. Mosallaei and Y. Rahmat-Samii, IEEE Trans. Antennas Propag. \textbf{49}, 60 (2001).

\bibitem{hsi}
H.-T. Chou, and Z.-D. Yan, IEEE Trans. Antennas Propag. \textbf{66}, 5794 (2018).

\bibitem{carl}
C. Pfeiffer and A. Grbic, IEEE Trans. Antennas Propag. \textbf{58}, 3055 (2010).

\bibitem{yuan}
Y. Su and Z. N. Chen, IEEE Trans. Antennas Propag. \textbf{66}, 5088 (2018).

\bibitem{soum}
S. Biswas and M. Mirotznik, Sci. Rep. \textbf{10}, 12646 (2020).

\bibitem{abda}
A. Dhouibi, S. N. Burokur, A. de Lustrac, and A. Priou, IEEE Antennas Wirel. Propag. Lett. \textbf{11}, 1504 (2012).

\bibitem{mark}
M. Bosiljevac, M. Casaletti, F. Caminita, Z. Sipus, and S. Maci, IEEE Trans. Antennas Propag. \textbf{60}, 4065 (2012).

\bibitem{dani}
D. Headland, W. Withayachumnankul, R. Yamada, M. Fujita, and T. Nagatsuma, APL Photonics \textbf{3}, 126105 (2018).

\bibitem{bohu}
B. Hu, T. Wu, Y. Cai, W. Zhang, and B.-L. Zhang, IEEE Access \textbf{8}, 4708 (2020).

\bibitem{torr1}
D. Torrent and J. S$\acute{a}$nchez-Dehesa, New J. Phys. \textbf{9}, 323 (2007).

\bibitem{peng}
S. Peng, Z. He, H. Jia, A. Zhang, C. Qiu, M. Ke, and Z. Liu, Appl. Phys. Lett. \textbf{96}, 263502 (2010).

\bibitem{zigo} L. Zigoneanu, B.-I. Popa, and S. A. Cummer, Phys. Rev. B \textbf{84}, 024305 (2011).

\bibitem{yong}
Y. Li, B. Liang, X. Tao, X.-F. Zhu, X.-Y. Zou, and J.-C. Cheng, Appl. Phys. Lett. \textbf{101}, 233508 (2012).

\bibitem{torr2}
D. Torrent, Y. Pennec, and B. Djafari-Rouhani, J. Appl. Phys. \textbf{116}, 224903 (2014).

\bibitem{jahd}
R. A. Jahdali and Y. Wu, Appl. Phys. Lett. \textbf{108}, 031902 (2016).

\bibitem{qian}
J. Qian, J.-P. Xia, H.-X. Sun, S.-Q. Yuan, Y. Ge, and X.-Z. Yu, J. Appl. Phys. \textbf{122}, 244501 (2017).

\bibitem{kim1}
S.-H. Kim, ``Sound Focusing by Acoustic Luneburg Lens," arXiv:1409.5489 (2014).

\bibitem{kim2}
S.-H. Kim, ``Cylindrical Acoustic Luneburg Lens," 8th International Congress on Advanced Electromagnetic Materials in Microwaves and Optics, (IEEE, 2014) pp.364-366.

\bibitem{park}
C. M. Park and S. H. Lee, Appl. Phys. Lett. \textbf{112}, 074101 (2018).

\bibitem{dong}
H. Y. Dong, Q. Cheng, G. Y. Song, W. X. Tang, J. Wang, and T. J. Cui, Appl. Phys. Express \textbf{10}, 087202 (2017).

\bibitem{yu}
L. Zhao and  M. Yu, J. Acoust. Soc. Am. \textbf{148}, EL82 (2020).

\bibitem{fang}
R. Zhu, C. Ma, B. Zheng, M. Y. Musa, L. Jing, Y. Yang, H. Wang, S. Dehdashti, N. X. Fang, and H. Chen, Appl. Phys. Lett. \textbf{110}, 113503 (2017).

\bibitem{fu-cummer}
Y. Fu, J. Li, Y. Xie, C. Shen, Y. Xu, H. Chen, and S. A. Cummer, Phys. Rev. Mater. \textbf{2}, 105202 (2018).

\bibitem{xie-cummer}
Y. Xie, Y. Fu, Z. Jia, J. Li, C. Shen, Y. Xu, H. Chen, and S. A. Cummer, Sci. Rep. \textbf{8}, 16188 (2018).

\bibitem{zhao}
L. Zhao, E. Laredo, O. Ryan, A. Yazdkhasti, H.-T. Kim, R. Ganye, T. Horiuchi, and M. Yu, Appl. Phys. Lett. \textbf{116}, 071902 (2020).

\bibitem{fokin}
V. Fokin, M. Ambati, C. Sun, and X. Zhang, Phys. Rev. B \textbf{76}, 144302 (2007).

\bibitem{boyl1}
C. A. Boyles, J. Acoust. Soc. Am. {\textbf4\textbf5}, 351 (1969).

\bibitem{boyl2}
C. A. Boyles, J. Acoust. Soc. Am. {\textbf4\textbf5}, 356 (1969).

\bibitem{miao}
Y. Chen, H. Liu, M. Reilly, H. Bae, and M. Yu, Nat. Commun. \textbf{5}, 5247 (2014).

\bibitem{tol}
S. Tol, F. L. Degertekin and A. Erturk, Appl. Phys. Lett. \textbf{109}, 063902 (2016).

\bibitem{jin}
Y. Jin, R. Kumar, O. Poncelet, O. Mondain-Monval, and T. Brunet, Nat. Commun. \textbf{10}, 143 (2019).

\bibitem{kim3}
S.-H. Kim, B.-W. Ahn, K.-M. Park, and G.-S. Lim, ``Sound Reception System by an Acoustic Luneburg Lens," arXiv:1906.07174 (2019).

\bibitem{cheng}
C. Ma, S. Gao, Y. Cheng, and X. Liu, Appl. Phys. Lett. \textbf{115}, 053501 (2019).

\end{thebibliography}
\end{document}